\documentclass{emulateapj}
\usepackage{graphicx}

\newcommand{\beq}{\begin{equation}}
\newcommand{\eeq}{\end{equation}}
\newcommand{\ba}{\begin{array}}
\newcommand{\ea}{\end{array}}

\newcommand{\ee}{\epsilon_{e,0}}

\def\be{\begin{equation}}
\def\ee{\end{equation}}

\begin{document}

\title{Dynamical Model of an Expanding Shell}

\author {Asaf Pe'er \altaffilmark{1}}
\altaffiltext{1}{Harvard-Smithsonian Center for Astrophysics, MS-51, 60
  Garden Street, Cambridge, MA 02138, USA}

\begin{abstract}
  Expanding blast waves are ubiquitous in many astronomical sources,
  such as supernovae remnants (SNRs), X-ray emitting binaries (XRBs)
  and gamma-ray bursts (GRBs).  I consider here the dynamics of such
  an expanding blast wave, both in the adiabatic and the radiative
  regimes. As the blast wave collects material from the surrounding,
  it decelerates. A full description of the temporal evolution of the
  blast wave requires consideration of both the energy density and the
  pressure of the shocked material. The obtained equation is different
  than earlier works in which only the energy was considered. The
  solution converges to the familiar results in both the
  ultra-relativistic and the sub-relativistic (Newtonian) regimes.
\end{abstract}





\section{Introduction}
\label{sec:intro}

Expanding blast waves are one of the most common phenomena in many
astronomical transients. Interaction of the expanding shells of
supernova remnants (SNRs) with their environment was studied more than
three decades ago \citep{Chevalier76, Chevalier82}. Similarly, in
X-ray emitting binaries (XRBs), mildly relativistic (Lorentz factor
$\Gamma \sim$~few) expanding radio blobs are observed for nearly two
decades now \citep{MiRo94, HR95, Fender+99}. In gamma-ray bursts
(GRBs) relativistic blast waves are an inherent part of the GRB
''fireball'' model \citep{Goodman86, Pac86, Shemi_Piran90, RM92,
  Narayan+92}, and are the source of the afterglow emission frequently
seen.

These expanding blast waves originate from a stellar explosion (such
as in GRBs or supernovae), or a rapid ejection of material (as in
XRBs). As they propagate through the ambient medium, they collect
material and decelerate. The expansion may be adiabatic, or highly
radiative. The dynamics of the blast wave expansion in both these
scenarios were extensively studied in the past \citep{BM76, KP97,
  CD99, Piran99, Huang+99, vPKW00}.\footnote{While \citet{BM76}
  provided a full solution in the ultra-relativistic, adiabatic limit,
  the original equation that describes the blast wave evolution in the
  radiative scenario was written and solved by \citet{KP97}. It was
  generalized to include both the radiative and adiabatic scenarios by
  \citet{CD99, Piran99}. However, it was later shown by
  \citet{Huang+99} that this equation does not converge to the
  correct asymptotic solution for adiabatic expansion in the Newtonian
  limit. \citet{Huang+99} thus modified this equation, solved the
  revised equation and showed that its solution converges to the
  correct asymptotic solutions in both the ultra-relativistic and the
  Newtonian limits, for both adiabatic and radiative
  scenarios. \citet{idiots1, idiots2, idiots3}, apparently being
  unaware of the work of \citet{Huang+99}, have re-solved the
  dynamical equation written by \citet{CD99, Piran99}, both numerically and
  analytically.  Apart from the work by \citet{BM76}, all the other
  works relied on similar basic assumptions, in which the contribution
  of the shocked-ISM pressure was neglected, and are therefore
  conceptually incorrect. In the following, I will compare the results
  derived here to the results derived by \citet{Huang+99}, that were
  shown to converge to the correct asymptotic limits.}  These
formulations are used until today as a basis for calculating the
expected radiation during the deceleration phase of the blast wave
evolution \citep[e.g.,][]{Narayan+11, GP12, VEM12, SM12}.

The dynamical calculations in the above mentioned works are based on a
``toy model'', in which a basic assumption is that the interaction
between the blast wave and the interstellar material (ISM) can be
described by a series of inelastic collisions \citep{KP97,
  Huang+99}. While this prescription asymptotes to the known
analytical solutions in the ultra-relativistic limit $\Gamma \gg 1$
\citep{MR97} and in the non-relativistic limit \citep{Sedov59}, it
neglects the contribution of the swept-up material to the internal
pressure. This, in turn, affects the evolution of the blast wave
dynamics. In this {\it letter}, I revise the basic assumptions of the
blast wave - ISM interaction scenario, and re-derive the equations
that govern the blast wave evolution. As will be shown below, the
obtained equations are different than the ones previously
used. Nonetheless, they asymptote to the known results at the
ultra-relativistic as well as the sub-relativistic limits.

\section{Dynamical model}
\label{sec:2}

Consider an explosion that ejects mass $M$ and creates a shell that
propagates into the cold ISM. In front of the expanding shell is a
blast wave. The system is thus composed of 3 regions: (1) the
unshocked ISM, (2) the shocked ISM, and (3) the ejected shell
material. For simplicity, I assume that a reverse shock wave (if
created) had long passed, hence the ejected shell material is
cold. Furthermore, I assume that the thermodynamical quantities of the
gas: $n_i, p_i$ and $e_i$ (particle number density, pressure and
internal energy density) are steady in each region. The pressure in
each region is given by $p' = ({\hat \gamma} -1)(e'-\rho')$, where
$\rho' = n' m_p c^2$ is the rest mass density and $\hat \gamma$ is the
adiabatic index. Here and below, primed quantities are in the comoving
frame, and unprimed quantities are in the observer's frame in which
the unshocked ISM is at rest.

The evolution of the blast wave as it propagates through the ISM is
calculated by considering energy conservation in the observer's frame.
Let us assume that at time $t$ the plasma propagates with Lorentz
factor $\Gamma$.\footnote{Note that the calculation here is general,
  and is not limited to ultra-relativistic speeds, where $\Gamma \gg
  1$.}  Neglecting radiative losses and assuming that the gas can be
described as a prefect fluid, the energy density in region (2) as is
measured by a distant observer is given by
\beq
\ba{lcl}
e_2 = T^{00} & = & (e'_2 + p'_2) u^0 u^0 + p'_2 g^{00} \nonumber \\
& =&  \left[\hat \gamma \Gamma^2 - (\hat \gamma - 1)\right] e'_2 +
(\hat \gamma -1) (1 - \Gamma^2) n'_2 m_p c^2.
\ea
\label{eq:1}
\eeq
Here, $T^{00}$ is the $00$ component of the stress-energy tensor. Due
to Lorentz contraction, the comoving volume of region (2) is
$V'=\Gamma V$. The total energy of the gas contained in this region
(in the adiabatic case), as is viewed by a distant observer is thus
\beq
\ba{lcl}
E_2(ad.) & = &  [\hat \gamma \Gamma^2 - (\hat \gamma - 1)] {e'_2 V' \over
  \Gamma} + (\hat \gamma -1)
(1 - \Gamma^2) {n'_2m_p c^2 V' \over \Gamma} \nonumber \\
& = &  [\hat \gamma \Gamma^2 - (\hat \gamma - 1)(1+\Gamma \beta^2)]
N_2 m_p c^2
\ea
\label{eq:2}
\eeq
where $N_2 = n_2 V$ is the number of particles in region (2), $\beta
\equiv (1 - \Gamma^{-2})^{1/2}$ is the normalized bulk velocity of the
plasma in this region, and use was made in the relation $e_2'/n_2' =
\Gamma m_p c^2$, which is exact for any value of $\Gamma$ as long as
the unshocked ISM in region (1) is cold \citep{BM76}.

The calculation in equation \ref{eq:2} assumes no radiative losses.
In order to allow the possibility that part of the thermal energy
gained by the ISM as it crosses the shock wave is radiated, equation
\ref{eq:2} is modified as follows.  The energy calculated in equation
\ref{eq:2} is the sum of three separate components: (1) the rest mass
energy of the shocked ISM; (2) its kinetic energy; and (3) its thermal
energy. The first two components sum up to $\Gamma N_2 m_p c^2$. Only the
energy in the third component can in principle be radiated. If a fraction
$\epsilon$ of the thermal energy is radiated, then the energy of the
gas in region (2) is given by
\beq
E_2 =  \left\{ \Gamma + (1-\epsilon)[\hat \gamma \Gamma^2 - \Gamma -
  (\hat \gamma - 1)(1+\Gamma \beta^2)]\right\} m c^2,
\label{eq:3}
\eeq
where $m \equiv N_2 m_p$ is the mass of the shocked ISM. 
The material in region (3) is assumed to be cold, and so its (kinetic
+ rest mass) energy is $E_3 = \Gamma M c^2$. 

A differential equation for the evolution of the plasma velocity is
derived in the following way.
Between times $t$ and $t+\delta t$, the plasma propagates a distance
$\beta c \delta t$, and an ISM of mass $dm$ crosses the forward shock
and gains kinetic and thermal energy. A fraction $\epsilon$ of the
gained thermal energy is assumed to be radiated, hence the radiated
energy is $\delta E_{rad} = \epsilon [\hat \gamma \Gamma^2 - \Gamma -
(\hat \gamma - 1)(1+\Gamma \beta^2)] dm c^2$. As it collects material,
the plasma decelerates; at time $t + \delta t$ its Lorentz factor is
$\Gamma - d \Gamma$ (corresponding velocity $\beta - d \beta$).
Conservation of energy at times $t$ and $t+ \delta t$ is written as 
\beq
\ba{lcl}
 & & \left\{ \Gamma + (1-\epsilon)[\hat \gamma \Gamma^2 - \Gamma -
  (\hat \gamma - 1)(1+\Gamma \beta^2)]\right\} m + \Gamma M 
\nonumber \\
& =&  -dm + \left\{ (\Gamma-d \Gamma) + (1-\epsilon)[\hat \gamma
  (\Gamma - d\Gamma)^2 - (\Gamma - d\Gamma)\right. \nonumber \\
& &\left. \quad \quad \quad -
  (\hat \gamma - 1)(1+(\Gamma-d\Gamma) (\beta-d\beta)^2)]\right\} (m
+ dm) \nonumber \\
& &+ (\Gamma - d\Gamma) M +  \epsilon [\hat \gamma (\Gamma -
d\Gamma)^2 - (\Gamma - d\Gamma) - \nonumber \\
& & \quad \quad \quad \quad \quad \quad \quad
(\hat \gamma - 1)(1+(\Gamma -d\Gamma) (\beta -d\beta)^2)] dm.
\ea
\label{eq:4a}
\eeq
Re-arranging the terms in equation \ref{eq:4a}, it can be written as
a dynamical equation for the evolution of the bulk motion Lorentz factor, 
\beq
{d\Gamma \over dm} = -{\hat \gamma(\Gamma^2 -1) - (\hat \gamma - 1)
  \Gamma \beta^2 \over M + \epsilon m + (1 - \epsilon) m [2 \hat
  \gamma \Gamma - (\hat \gamma -1) ( 1+\Gamma^{-2})]}.
\label{eq:4}
\eeq
Equation \ref{eq:4} thus describes the evolution of the plasma Lorentz
factor due to its interaction with the ISM. This equation holds both
in the ultra-relativistic ($\Gamma \gg 1$) as well as the
sub-relativistic ($\beta \ll 1$) limits. It can be compared to
equation (7) in \citet{Huang+99}, which, as described above, was
derived based on the assumption of continuous inelastic collisions
between the blast wave and the ISM, and hence does not contain the
contribution of the shock-heated ISM to the pressure in region
(2).\footnote{Note that equation (7) of \citet{Huang+99} is obtained
  by setting $\hat \gamma = 1$ in equation \ref{eq:4}.}

\subsection{Asymptotic limits}
\label{sec:analytic}

It is possible to obtain analytical solutions to the dynamical equation
\ref{eq:4} in the limits of ultra-relativistic and sub-relativistic
limits. These are useful for comparison with former results.

In the {\it adiabatic} scenario, $\epsilon = 0$, and equation
\ref{eq:4} reduces to
\beq
{d\Gamma \over dm} = -{\hat \gamma(\Gamma^2 -1) - (\hat \gamma - 1)
  \Gamma \beta^2 \over M +  m [2 \hat
  \gamma \Gamma - (\hat \gamma -1) ( 1+\Gamma^{-2})]}.
\label{eq:5}
\eeq
In the ultra-relativistic limit, $\Gamma \gg 1$, equation
\ref{eq:5} can be re-written as 
\beq
{d\Gamma \over dm} \simeq -{(\Gamma^2 -1) \over (M /\hat \gamma) +  2
\Gamma m }.
\label{eq:6}
\eeq
Denoting by $\Gamma_0$ the initial Lorentz factor of the flow, the
evolution of the plasma Lorentz factor can be separated into two
regimes. (1) Initially, $M/\hat \gamma \gg \Gamma_0 m$. In this
regime, $\Gamma\simeq \Gamma_0$.  (2) For $\Gamma m \gg M/\hat \gamma$
one obtains the familiar solution, $\Gamma \propto m^{-1/2}$, which,
for constant density ISM ($n \propto r^0$, $m \propto r^3$) results in
the well known solution $\Gamma \propto r^{-3/2}$. \footnote{Equation
  \ref{eq:6} does not admit a third regime, $\Gamma_0 m \gg M \gg
  \Gamma m$.}  Interestingly, equation \ref{eq:6} is similar to
equation 8 of \citet{Huang+99}, with $M$ replaced by $M/\hat \gamma
\simeq (3/4)M$. This discrepancy has only a minor effect on the blast
wave evolution.

On the other extreme, the sub-relativistic limit $\beta \ll 1$,
equation \ref{eq:5} reduces to 
\beq
{d (\Gamma \beta) \over dm} \simeq -{\beta \over M +  m [2 -\beta^2]},
\label{eq:7}
\eeq
which admits the solution $\beta \propto m^{-1/2}$, for $m \gg
M$. Thus, the general solution describing the blast wave evolution in
the adiabatic scenario (equation \ref{eq:5}) for $\Gamma m \gg M/\hat
\gamma$ can be written as $\Gamma \beta \propto m^{-1/2}$.

In the {\it radiative} scenario, $\epsilon = 1$. In the
ultra-relativistic limit $\Gamma \gg 1$, the dynamical equation
(eq. \ref{eq:4}) becomes
\beq
 {d \Gamma \over dm} \simeq -{\hat
  \gamma \Gamma^2 \over M + m}.
\label{eq:8}
\eeq
Initially, $M \gg m$, resulting in a steady Lorentz factor, $\Gamma
\simeq \Gamma_0$. At a later stage, $\Gamma_0 m \gg M$ and one
obtains the decay law $\Gamma(m) \simeq M/\hat \gamma m$ (as long as
$m \ll M$). For
constant density ISM, $n \propto r^0$, this leads to the familiar
decay law, $\Gamma \propto r^{-3}$.  This result is different than the
result of \citet{Huang+99} by a factor $(\hat \gamma)^{-1} \simeq
3/4$.

In the sub-relativistic limit, $\beta \ll 1$, equation \ref{eq:4}
becomes 
\beq
{d (\Gamma \beta) \over dm} \simeq -{\beta \over M +  m},
\label{eq:9}
\eeq
with the solution $\beta \propto m^{-1}$ (for $m \gg M$). This
solution is similar to the classical ``snowplow'' evolution of an
expanding supernova remnants in the radiative regime
\citep{Spitzer68}.

\section{Numerical solution}
\label{sec:results}

A full solution of the dynamic equation \ref{eq:4} can easily be
obtained numerically. In solving this equation, one first needs to
determine the value of the adiabatic index $\hat \gamma$, which
depends on the gas temperature. In calculating $\hat \gamma$, I assume
that the gas in region (2) maintains a Maxwellian distribution with
normalized temperature $\theta \equiv k_B T / m_p c^2$.  The average
energy per particle in this region is thus given by $<e'_2/n'_2 m_p
c^2> = K_1(\theta^{-1})/K_2(\theta^{-1}) + 3 \theta$. Here, $K_1, K_2$
are modified Bessel $K$-functions of the second kind \citep[][
5.34]{Lightman+75}.

The ratio $<e'_2/n'_2 m_p c^2 > = \Gamma$ is determined by the shock
jump conditions, and is thus known at any given instance. For a given
$\Gamma$, a good fit to the normalized temperature is
\beq 
\theta \simeq \left({\Gamma \beta \over 3}\right) \left({\Gamma \beta +1.07 (\Gamma
  \beta)^2 \over 1+\Gamma \beta+1.07 (\Gamma\beta)^2}\right).
\label{eq:10}
\eeq
This fit asymptotes to the exact solution in the limits $\Gamma \gg 1
$ and $\beta \ll 1$. The maximum error found is less than $3\times
10^{-3}$, for $\Gamma \beta \simeq 1$.  Once the gas temperature is
calculated, I use the polynomial fit given by \citet{Service86}, to
calculate $\hat \gamma \simeq
(5-1.21937z+0.18203z^2-0.96583z^3+2.32513z^4-2.39332z^5+1.07136z^6)/3$,
where $z \equiv \theta/(0.24+\theta)$. This fit is accurate to
$10^{-3}$.

Equation \ref{eq:4} is solved using 4th order Runge-Kutta method.  I
consider two sets of parameters, one representing GRB and one XRB. In
both scenarios, the blast wave is assumed to propagate into a
constant density ISM, and hence the collected mass is
\beq
dm  = 4 \pi r^2 n m_p dR,
\label{eq:11}
\eeq
where $n$ is the number density of the ISM and $m_p$ is the proton
rest mass. Photons emitted as the plasma propagates a distance $dR$
are observed at time $dt$, given by
\beq
dR = \Gamma \beta c \left( \Gamma + \Gamma \beta\right) dt.
\label{eq:12}
\eeq
While equation \ref{eq:12} is derived under the assumption that the
observed photons are emitted from a plasma which propagates towards
the observer, a more comprehensive calculation which considers the
integrated emission from different angles to the line of sight results
in a similar solution, up to a numerical factor of a few
\citep{Waxman97, PW06}.

\begin{figure}
\plotone{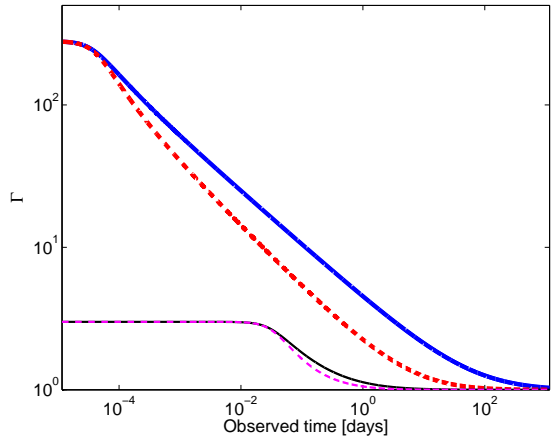}
\caption{Temporal evolution of the bulk Lorentz factor,
  $\Gamma$. Thick lines are for ``GRB'' scenario ($\Gamma_0 = 278$, $M
  = 4 \times 10^{28}$~gr), while thin lines are for ``XRB'' scenario
  ($\Gamma_0 = 3$, $M = 3 \times 10^{23}$~gr). In all cases, ISM
  density $n=1 {\rm \, cm^{-3}}$ assumed. Solid: adiabatic, $\epsilon=0$ (blue,
  black: GRB, XRB respectively); Dashed: radiative, $\epsilon=1$ (red, magenta: GRB,
  XRB respectively). }
\label{fig:1}
\end{figure}

\begin{figure}
\plotone{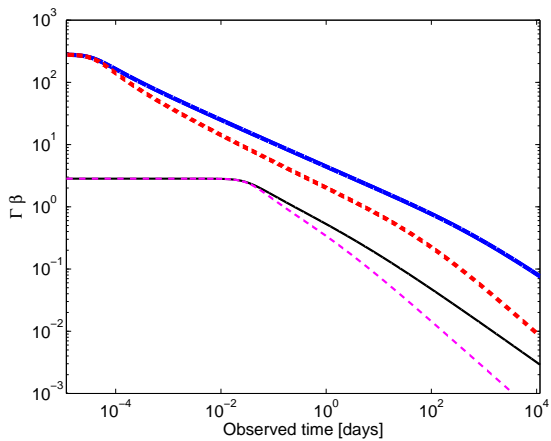}
\caption{Temporal evolution of the momentum, $\Gamma \beta$. All
  parameters values are similar to Figure \ref{fig:1}. }
\label{fig:2}
\end{figure}

\begin{figure}
\plotone{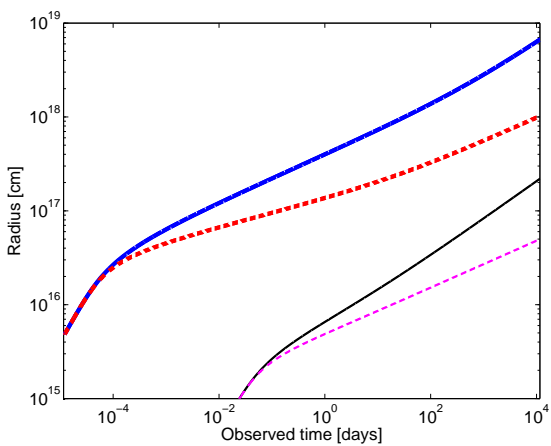}
\caption{Temporal evolution of the shock radius, $R$.  All
  parameters values are similar to Figure \ref{fig:1}.  }
\label{fig:3}
\end{figure}

The evolution of the Lorentz factor $\Gamma$, the momentum $\Gamma
\beta$ and the radius $R$ are presented in figures \ref{fig:1} --
\ref{fig:3}. In solving the dynamical equation, an ISM density of $n=1
{\rm \, cm^{-3}}$ is taken. For the initial explosion conditions, two
sets of parameters are used. The first set is representative for GRBs:
I take $E = 10^{52}$~erg and $M = 2 \times 10^{-5} M_\odot$, resulting
in $\Gamma_0 = 278$ \citep{Huang+99}.  The second set is
representative for XRBs, and is chosen as follows. As an initial
Lorentz factor I take a fiducial value $\Gamma_0 = 3$
\citep{Miller-Jones+05}.  Observed XRB radio blobs are typically
emitted when the luminosity is close to the Eddington luminosity, and
the flux rise time is a few days, hence the total energy released is
of the order of $\sim 10^{45}$~erg \citep[e.g.,][]{FBG04}.  With
$\Gamma_0 = 3$ this leads to an ejected mass of $M = 3 \times
10^{23}$~gr.

The results in Figures \ref{fig:1} -- \ref{fig:3} are given for both
the adiabatic scenario and the radiative scenario. As expected, the
results asymptote to the known solutions, which can be divided into 3
regimes. (I) Initially, $\Gamma \simeq \Gamma_0$, and $R\propto t$;
(II) $\Gamma m \gg M$, and $\Gamma \gg 1$. In the {\it adiabatic}
scenario, this leads to $R(t) \propto t^{1/4}$ and $\Gamma(t) \propto
t^{-3/8}$, while in the {\it radiative} scenario, $R(t) \propto
t^{1/7}$ and $\Gamma(t) \propto t^{-3/7}$. (III) For $\beta \ll 1$, in
the {\it adiabatic} scenario $R(t) \propto t^{2/5}$ and $\beta(t)
\propto t^{-3/5}$, while in the {\it radiative} scenario, $R(t)
\propto t^{1/4}$ and $\beta(t) \propto t^{-3/4}$.

\section{Summary and discussion}
\label{sec:summary}

In this {\it letter}, I have revisited the dynamics of a blast wave
propagating into an ISM, as is expected in many astronomical objects,
such as GRBs, XRBs and Supernovae. I derived an equation (\ref{eq:4})
that determines the evolution of the blast wave Lorentz factor as the
plasma collects material from the ISM and decelerates. Analytical
solutions in both the adiabatic and the radiative regimes are found in
the asymptotic limits $\Gamma \gg 1$ and $\beta \ll 1$
(\S\ref{sec:analytic}). Numerical integration is easily carried, and
the resulting dynamics valid in the full regime is presented 
\S\ref{sec:results}. 

The dynamical equation is different than the dynamical equations
derived earlier by several authors \citep{KP97,Huang+99, CD99,
  Piran99}. This is due to a conceptual difference: earlier works
considered a ``toy model'', in which the basic assumption is that the
interaction between the blast wave and the interstellar material (ISM)
can be described by a series of inelastic collisions. As opposed to
that, here I consider the full energy-momentum tensor, which takes
into account the contribution of the collected material to both the
energy and the pressure in the shocked region. Such an inclusion was
neglected in earlier works.

While the results of earlier works retrieve the correct asymptotic
behavior in the limits $\Gamma \gg 1$ and $\beta \ll 1$
\citep{Huang+99}, the evolution of the Lorentz factor derived in these
works is different than the evolution derived here by a numerical
factor of tens of percents.  In the {\it radiative} scenario, the
difference in the derived value of the Lorentz factor at a given
radius is up to $\gtrsim 30\%$. This result is not surprising, given
that the main difference between the dynamics derived here and the
dynamics derived in former works lies in the inclusion of $\hat \gamma
\simeq 4/3$ (for $\Gamma \gg 1$). Interestingly, also in the {\it
  adiabatic} scenario the numerical difference is larger than $12 \%$.

In recent years, a renewed interest in calculating the expected flux
from the interaction of an expanding blast wave with its environment
had emerged. Works had been carried in the context of GRB afterglow
emission \citep{GE10, GP12, VEM12, Xu+12}, the evolution of the
observed radio blobs seen in XRBs \citep{SM12, NM12} as well as the
evolution of emission from supernovae \citep{ChR11}. The dynamical
calculations presented here are an obvious essential step in these
calculations.

\acknowledgments
I would like to thank Ramesh Narayan, Lorenzo Sironi and Ralph Wijers for useful discussions.


\bibliographystyle{/Users/apeer/Documents/Bib/apj}



\end{document}